\documentclass[11pt,twoside,onecolumn]{article}
\usepackage[]{latexsym}
\usepackage{epsfig}
\usepackage{amsmath,amssymb}
\RequirePackage[dvipsnames,usenames]{color}
\newcommand{\be}{\begin{eqnarray}}
\newcommand{\ee}{\end{eqnarray}}

\title{\bf Using MGD gravitational decoupling to extend the isotropic solutions of Einstein equations to the anisotropical domain}
\author{C. las Heras\thanks{camilo.lasheras@ua.cl}
$\,$ 
P. Leon\thanks{pablo.leon@ua.cl}
$\,$ 
\null
\\
$^c${\em Departamento de F\'isica, Universidad de Antofagasta}
\\
\\
{\em  Antofagasta, Chile}
}
\begin{document}
\maketitle
\begin{abstract}
The aim of this work is to obtain new analitical solutions for Einstein equations in the anisotropical domain. This will be done via the minimal geometric deformation (MGD) approach, that allow us to decouple the Einstein equations. It requires a perfect fluid known solution that we will choose to be Finch-Skeas(FS) solution. Two different constraints were applied, and in each case we found an interval of values for the free parameters, where necesarly other physical solutions shall live.


\end{abstract}
%
%
%
%
%
%
%

\section{Introduction}
\setcounter{equation}{0}

One of the main features of the General Relativity (GR), is the relation between the space-time geometry and the energy-matter content of the universe. This relation appears in Einstein's Equations, which is a set of coupled nonlinear differential equations. 

A problem for theoretical physics and mathemathics is to obtain analytical solutions from these set of equations. This has been done only for specific distributions as the isotropic perfect fluid \cite{Stephani}. However,  there are reasons to think about the existence of anisotropy in more realistic compact objects \cite{Luis,herre1,herre2,herre3,angel}.
In that sense, it is important to obtain analytical solutions of Einstein's equations for a more general energy-momentum tensor that describes anisotropy in the pressure.

A method has been found in recent years, that is able to obtain solutions of the Einstein equations in the context of the Randall-Sundrum brane-world \cite{jo1,jo2}. This method is known as the  Minimal Geometric Deformation (MGD) approach, and it has a huge range of applications \cite{MGDextended1,MGDextended2,so,jo6,jo8,jo9,jo10,jo11,jo12,roldaoGL,rrplb,rr-glueball,rr-acustic,casa}. Among these applications, there is one of particular interest that allow us to decouple gravitational sources in GR \cite{MGD-decoupling,MGD-2}. In consequence we are able to find analytical internal solutions of Einstein equations that considers an energy-momentum tensor of the form
\begin{eqnarray}\label{1.1}
T_{\mu\nu}=T^{(PF)}_{\mu\nu}+\alpha\theta_{\mu\nu},
\end{eqnarray}
where $\alpha$ is a coupling constant and $\theta_{\mu\nu}$ is a gravitational source. In fact, with this approach we are able to study different systems as polytropic spheres \cite{zdenek1,zdenek2,zdenek3}, Horava-aether gravity \cite{carloni2017,ted}, Einstein-Maxwell \cite{EM}, Einstein Klein-Gordon \cite{shapiro,bosonstars,konstantin,singleton}, and many others (see for example \cite{diFelice,thomas,salvatore,salvatore2,marcelo1,marcelo2,Alvarez,stelle,sengupta,roman}).  

This works for as many gravitational sources as we want
\begin{eqnarray}
T^{(n)}_{\mu\nu}=T^{(n-1)}_{\mu\nu} + \alpha^{(n)}T^{(n)}_{\mu\nu} \hspace{0.5cm}\mbox{     with       } \hspace{0.5cm} T^{(0)}_{\mu\nu}=T^{PF}_{\mu\nu}
\end{eqnarray}
as long the spherically symmetry is preserved and the interaction between the sources is purely gravitational.
\begin{eqnarray}
\nabla_\nu T^{\mu\nu}=\nabla_\nu T^{(1)\mu\nu}= \dots = \nabla_\nu T^{(n)\mu\nu}=0
\end{eqnarray}

The first step in this new method is to choose two different gravitational sources. After that, we must solve the standard Eistein equations for one of them (usually the simplest). Then we have to deal we a set of equations similar to the Einstein equations system related to the other source. Finally with the combination of the two solutions, we will obtain the final solution for a system composed  by the two sources. 

In this paper we will study the Einstein equations with an energy-momentum tensor that considers anisotropy induced in general by the source. The decoupling of the equations with the MGD method and the matching conditions between the interior and the expected exterior solution will be explained in section 3. We will then apply the MGD method in section 4 to the well known FS solution \cite{FS}. Then we will obtain an interval of the coupling constants where physical solutions shall live considering the mimic constraint for pressure and density. Indeed we found new analytic and physical solutions specified in each case. Finally we summarize our conclusions in section 5.

\section{Einstein equations}
\label{s2}
\setcounter{equation}{0}
%
%

If we consider a static spherically symmetric distribution whose line element can be written in Schwarzschild-like coordinates as
\begin{equation}
ds^{2}
=
e^{\nu (r)}\,dt^{2}-e^{\lambda (r)}\,dr^{2}
-r^{2}\left( d\theta^{2}+\sin ^{2}\theta \,d\phi ^{2}\right)
\ ,
\label{metric}
\end{equation}
where $\nu =\nu (r)$ and $\lambda =\lambda (r)$ are functions of the areal
radius $r$ only. Then this metric must satisfy the Einstein equations
\begin{equation}
\label{corr2}
R_{\mu\nu}-\frac{1}{2}\,R\, g_{\mu\nu}
=
-8\pi\,T^._{\mu\nu}.
\ 
\end{equation}

Now, if we choose an energy-momentum tensor of the form
\begin{equation}
\label{emt}
T_{\mu\nu}
=
T^{\rm (PF)}_{\mu\nu}+\alpha\,\theta_{\mu\nu}
\ ,
\end{equation}
where 
\begin{equation}
\label{perfect}
T^{\rm (PF)}_{\mu \nu }=(\rho +p)\,u_{\mu }\,u_{\nu }-p\,g_{\mu \nu },
\end{equation}
is the  matter-energy content of a perfect fluid with $u^{\mu }=e^{-\nu /2}\,\delta _{0}^{\mu }$ the fluid 4-velocity. Then the Einstein equations leads to 
\begin{eqnarray}
\label{ec1}
8\pi
\underbrace{\left(
\rho+\alpha\,\theta_0^{\,0}
\right)}_{\tilde{\rho}}
&\!\!=\!\!&
\frac 1{r^2}
-
e^{-\lambda }\left( \frac1{r^2}-\frac{\lambda'}r\right)\ ,
\\
\label{ec2}
8\pi
\underbrace{\left(p-\alpha\,\theta_1^{\,1}\right)}_{\tilde{p_r}}
&\!\!=\!\!&
-\frac 1{r^2}
+
e^{-\lambda }\left( \frac 1{r^2}+\frac{\nu'}r\right)\ ,
\\
\label{ec3}
8\pi
\underbrace{\left(p-\alpha\,\theta_2^{\,2}\right)}_{\tilde{p_t}}
&\!\!=\!\!&
\frac {e^{-\lambda }}{4}
\left( 2\,\nu''+\nu'^2-\lambda'\,\nu'
+2\,\frac{\nu'-\lambda'}r\right)
\ ,
\end{eqnarray}
where the prime indicates derivatives respect to variable $r$, ranging from the star center ($r=0$) to the surface of the star  ($r=R$).

The total energy momentum tensor (\ref{emt}) must satisfy the conservation equation
\begin{equation}
\label{con1}
\nabla_\nu T^{\mu\nu}=
p'
+
\frac{\nu'}{2}\left(\rho+p\right)
-
\alpha\left(\theta_1^{\,\,1}\right)'
+
\frac{\nu'}{2}\alpha\left(\theta_0^{\,\,0}-\theta_1^{\,\,1}\right)
+
\frac{2\,\alpha}{r}\left(\theta_2^{\,\,2}-\theta_1^{\,\,1}\right)
=
0
\ ,
\end{equation}
where the perfect fluid case is recovered for $\alpha= 0$.

After identifying an effective density $\tilde{\rho}$, an effective isotropic pressure $\tilde{p_r}$ and an effective tangential pressure $\tilde{p_t}$, the system of eqs (\ref{ec1})-(\ref{ec3}) can be related to an anisotropical one \cite{Luis,tiberiu} where the unknown functions are ($\tilde{\rho}$,$\tilde{p_r}$,$\tilde{p_t}$,$\lambda$ and $\nu$). 

The anisotropy is given by
\begin{equation}
\label{anisotropy}
\Pi
\equiv
\tilde{p}_{t}-\tilde{p}_{r}
=
\alpha\left(\theta_1^{\,1}-\theta_2^{\,2}\right)
\ ,
\end{equation}
inside the stellar distribution.

In order to solve the system, we need to get more information than the one given by the energy-momentum tensor. Considering this, we must establish a relation between the physical variables of the system. This can be done via a state equation, or as it is in our case via the MGD approach.

\section{MGD Approach and Matching Conditions}
\setcounter{equation}{0}
\label{s3}

The first step in the MGD approach is to take a general relativistic perfect fluid solution of Einstein equations ($\alpha=0$), with a line element given in Schwarzschild-like coordinates by 
\begin{equation}
ds^{2}
=
e^{\Xi (r)}\,dt^{2}
-
\frac{dr^{2}}{\mu(r)}
-
r^{2}\left( d\theta^{2}+\sin ^{2}\theta \,d\phi ^{2}\right)
\ ,
\label{pfmetric}
\end{equation}
where 
\begin{equation}
\label{standardGR}
\mu(r)
\equiv
1-\frac{8\pi}{r}\int_0^r x^2\,\rho\, dx
=1-\frac{2\,m(r)}{r},
\end{equation}
is the standard expression for the mass function in GR. 

Now, in order to consider the anisotropy generated by the gravitational source $\theta_{\mu\nu}$ in the equations (\ref{ec1})-(\ref{ec3}), we have to include the parameter $\alpha$ in the perfect fluid solution (\ref{pfmetric}). One way to do that is to assume that the effects due to the $\alpha$ parameter are included in the deformation of the geometric functions $\Xi$ and $\mu$.
\begin{eqnarray}
\label{gd1}
\Xi
&\mapsto &
\nu
=
\Xi+\alpha\,g
\ ,
\\
\label{gd2}
\mu 
&\mapsto &
e^{-\lambda}
=
\mu+\alpha\,f
\ ,
\end{eqnarray}
where $g$ and $f$ are the deformations related to the temporal and radial
metric component, respectively. In the particular case where $g=0$, we have
\begin{eqnarray}
\label{expectg}
\mu(r)\mapsto\,e^{-\lambda(r)}
=
\mu(r)+\alpha\,f^{*}(r)
\ ,
\end{eqnarray}
which is known as the Minimal Geometric Deformation.

Using now this expression is easy to see that the Einstein equations~(\ref{ec1})-(\ref{ec3}) splits in two systems. The first of them is
\begin{eqnarray}
\label{ec1pf}
8\pi\rho
&\!\!=\!\!&
\frac{1}{r^2}
-\frac{\mu}{r^2}
-\frac{\mu'}{r}\ ,
\\
\label{ec2pf}
8\pi\,p
&\!\!=\!\!&
-\frac 1{r^2}+\mu\left( \frac 1{r^2}+\frac{\nu'}r\right)\ ,
\\
\label{ec3pf}
8\pi\,p
&\!\!=\!\!&
\frac{\mu}{4}\left(2\nu''+\nu'^2+\frac{2\nu'}{r}\right)+\frac{\mu'}{4}\left(\nu'+\frac{2}{r}\right)
\ ,
\end{eqnarray}
which turns out to be the perfect fluid Einstein equations, and the associated conservation equation is (\ref{con1}) with $\alpha=0$
\begin{eqnarray}
\label{conpf}
p'+\frac{\nu'}{2}\left(\rho+p\right) = 0.
\end{eqnarray}

On the other hand, the second system of equations reads
\begin{eqnarray}
\label{ec1d}
8\pi\,\theta_0^{\,0}
&\!\!=\!\!&
-\frac{f^{*}}{r^2}
-\frac{f^{*'}}{r}
\ ,
\\
\label{ec2d}
8\pi\,\theta_1^{\,1}
&\!\!=\!\!&
-f^{*}\left(\frac{1}{r^2}+\frac{\nu'}{r}\right)
\ ,
\\
\label{ec3d}
8\pi\,\theta_2^{\,2}
&\!\!=\!\!&
-\frac{f^{*}}{4}\left(2\nu''+\nu'^2+2\frac{\nu'}{r}\right)
-\frac{f^{*'}}{4}\left(\nu'+\frac{2}{r}\right)
\ .
\end{eqnarray}
and the conservation equation associated with the source is
\begin{eqnarray}
\label{con1d}
\left(\theta_1^{\,\,1}\right)'
-\frac{\nu'}{2}\left(\theta_0^{\,\,0}-\theta_1^{\,\,1}\right)
-\frac{2}{r}\left(\theta_2^{\,\,2}-\theta_1^{\,\,1}\right)
=
0.
\end{eqnarray}

In this case, Eqs (\ref{ec1d})-(\ref{ec3d}) is not an Einstein equations system due to a missing term $\frac{1}{r}$ on the right side of the first two equations.

However, is easy to see that if we define an energy-momentum tensor $\tilde{\theta}_{\mu\nu}$ given by
\begin{eqnarray}
\label{theta1}
\tilde{\rho}
&\!\!=\!\!&
{\theta^*}_0^{\,\,0}=\theta_0^{\,0}+\frac{1}{k^2\,r^2}
\ ,
\\
\label{theta2}
\tilde{p}_r
&\!\!=\!\!&
{\theta^*}_1^{\,\,1}=\theta_1^{\,1}+\frac{1}{k^2\,r^2}
\ ,
\\
\label{theta3}
\tilde{p}_t
&\!\!=\!\!&
{\theta^*}_2^{\,\,2}=\theta_2^{\,2}={\theta^*}_3^{\,\,3}=\theta_3^{\,3}
\ ,
\end{eqnarray}
the system of equations (\ref{ec1d})-(\ref{ec3d}) is equivalent to Einstein equations for an anisotropic fluid with energy-momentum tensor $\tilde{\theta}_{\mu\nu}$.

Furthermore, the conservation equations (\ref{conpf}) y (\ref{con1d}) implies that the interaction of the perfect fluid with the source is purely gravitational.

Finally, in order to avoid the appearance of singular behaviour of the physical variables on the surface of our distribution, we must impose matching conditions between the interior and the exterior space-time geometries.

The inner region is defined by the MGD metric
\begin{equation}
ds^{2}
=
e^{\nu^{-}(r)}\,dt^{2}
-\left(1-\frac{2\,\tilde{m}(r)}{r}\right)^{-1}dr^2
-r^{2}\left(d\theta ^{2}+\sin {}^{2}\theta d\phi ^{2}\right)
\ ,
\label{mgdmetric}
\end{equation}

where the interior mass function is given by
\begin{equation}
\label{effecmass}
\tilde{m}(r)
=
m(r)-\frac{r}{2}\,\alpha\,f^{*}(r)
\ , 
\end{equation}
with $m$ given by the standard GR expression in Eq.~(\ref{standardGR}).

The most general outer region is the one where there is no perfect fluid. In majority of the cases it will not be an exterior Swcharzschild solution due to the contribution of $\tilde{\theta}_{\mu\nu}$.

If we consider the particular case in which the outher region is described by the the exterior Schwarzschild solution, it can be shown \cite{MGD-2} that the first and second fundamental form are given by
\begin{eqnarray}
\Big(1-\frac{2\,\cal M}{r}+\alpha\,f^{*}_{r}
\Big)_{R^-}&=&
\Big(1-\frac{2{\cal M}}{r}\Big)_{R^+}
\ ,
\label{ffgeneric1} \\
\tilde{p}_R\equiv\,p_{R}+\alpha\,\frac{f_{R}^{\ast }}{k^2}
\left(\frac{1}{R^{2}}+\frac{\nu _{R}^{\prime }}{R}\right)&=&0
\ .
\label{pnegative}
\end{eqnarray}
where $\mathcal{M}=m(R)$ and $p_{R}\equiv p^{-}(R)$.

\section{New exact anisotropic solutions of Einstein equations}
\label{s5}
\setcounter{equation}{0}
In order to apply the MGD approach, the perfect fluid solution with physical relevance that we have chosen is the well-known FS solution $\{\nu,\mu,\rho, p\}$ \cite{FS}, namely 
\begin{eqnarray}
e^{\nu(r)} &=& D^2 [(B-A\xi)\cos(\xi)+(A+B\xi)\sin(\xi)]^2 \label{tolman00} , \\
\mu(r) &=& \xi^2, \label{tolman11} \\
\frac{8\pi \rho(r)}{C} &=& \frac{2+\xi ^2}{\xi^4}, \label{tolmandensity} \\
\frac{8\pi p(r)}{C} &=& -\frac{1}{\xi^2}\left(\frac{(\beta \xi +1)+(\beta - \xi)\tan(\xi)}{(\beta \xi -1)-(\beta + \xi)\tan(\xi)}\right), \label{tolmanpressure}
\end{eqnarray}
where $\xi=\sqrt{1+Cr^2}$, $\beta=A/B$, $D$ is a scaling constant that can take any value. 
From (\ref{tolmanpressure}), we define the surface as the value of $\xi$ ($\xi_s$) where $p(r_s)=0$, this implies that 
\begin{equation}
\label{A}
\beta
=
\frac{\xi_s tan(\xi_s)-1}{tan(\xi_s)+1}.
\end{equation}
Now, from the matching condition between the interior and exterior solution the constant $C$ is given by
\begin{equation}
\label{B}
C
= \frac{2M}{r^2_s(r_s-2M)}
\end{equation}
This expression does not have to be the same when we consider the gravitational source $\theta_{\mu\nu}$.

If we consider $\alpha\neq 0$ in the interior,
using MGD approach as we describe it before, implies that we need more information to solve the system of equations. After that we have to make sure the physical acceptability conditions are satisfied in order to obtain an analytical solution with physical relevance in the anisotropical domain.

%
%
%
%
%
%
%
\subsection{Solution~I: A condition for the pressure}
\label{s5.1}

We choose to impose restrictions on the components of $\theta_{\mu\nu}$. Then it can be verified from the matching conditions between the interior and exterior solution, that $\theta_1^{\,\,1}(r)$ must be proportional to
$p(r)$. If we consider the simplest case we have \begin{equation}
\label{constraint}
\theta_1^{\,\,1}(r)
=
p(r)
\ .
\end{equation}

We notice from Eq.~(\ref{ec2d}) that this constraint (\ref{constraint}) implies that the radial metric component (\ref{expectg})  of the FS solution minimally deformed is given by  
\begin{equation}
\label{tolman11d}
e^{-\lambda(r)}
= 
\xi^2-\alpha \left(\frac{\xi^2-1}{\xi^2}\right)\frac{H}{1+H(\xi^2-1)},
\end{equation}
where $\displaystyle H=\frac{8\pi p}{C}$.

It can be seen that the limit $\alpha\to 0$ leads to the standard F-S
solution for perfect fluids. The temporal metric component of the FS solution (\ref{tolman00}), it does not change in the MGD approach due to the minimally character of the deformation. 

\begin{figure}[h!]
\begin{center}
\includegraphics[scale=0.4]{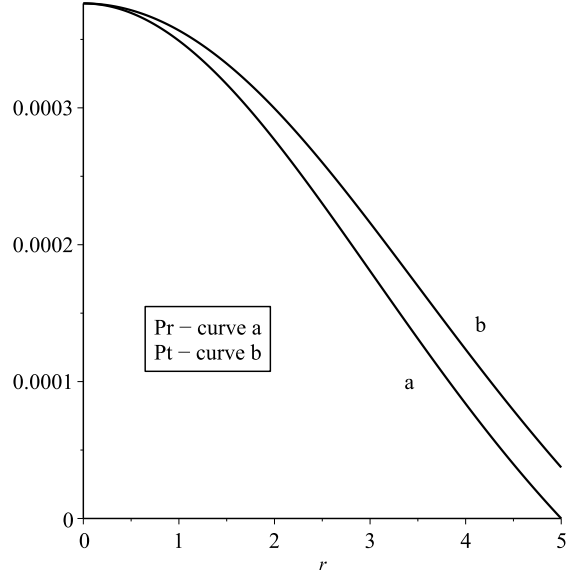}
\end{center}
\caption{Radial and tangential pressure}
\label{prpt02}
\end{figure} 
\begin{figure}[h!]
\begin{center}
\includegraphics[scale=0.4]{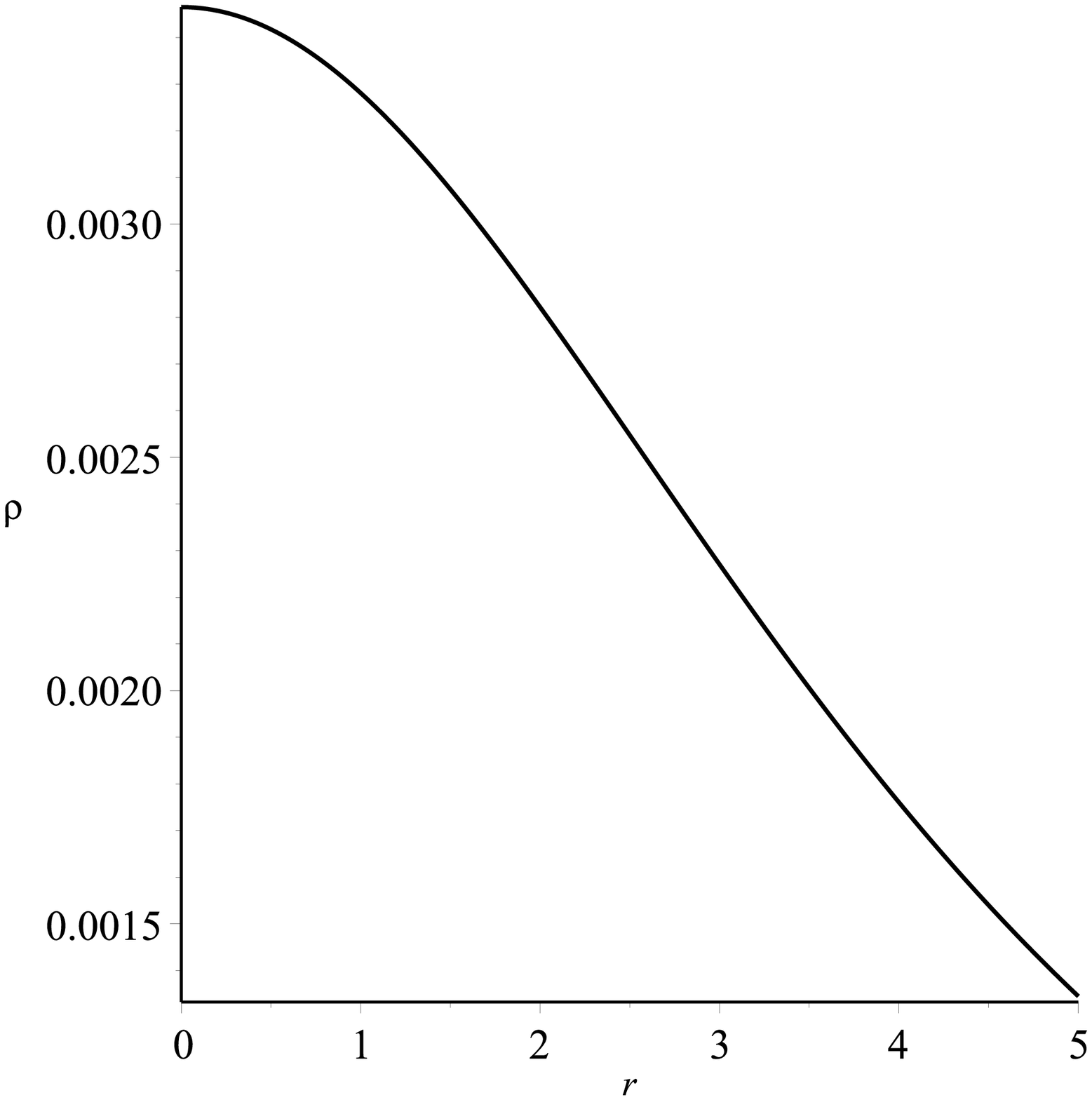}
\end{center}
\caption{Density.}
\label{rho02}
\end{figure} 
\begin{figure}[h!]
\begin{center}
\includegraphics[scale=0.4]{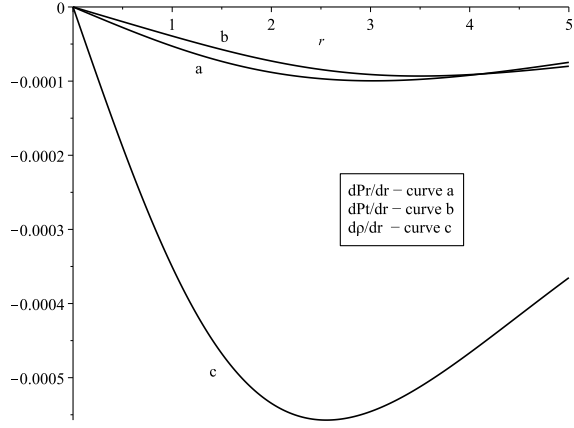}
\end{center}
\caption{Gradient for Radial pressure, tangential pressure, and density.}
\label{derivadasr02}
\end{figure}
\begin{figure}[h!]
\begin{center}
\includegraphics[scale=0.4]{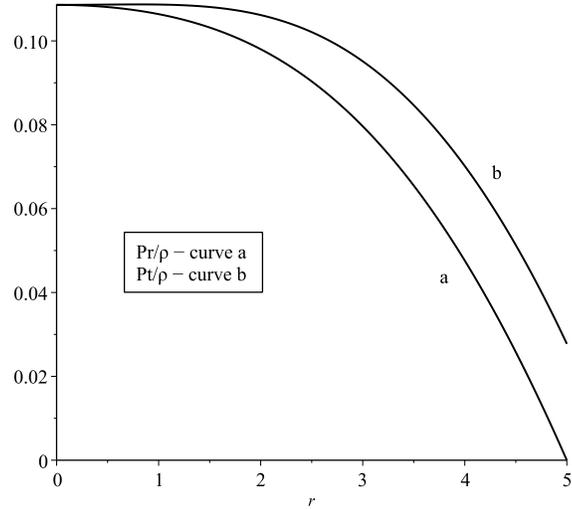}
\end{center}
\caption{Quotients of radial pressure, tangential pressure, and density.}
\label{coientesP02}
\end{figure}
\begin{figure}[h!]
\begin{center}
\includegraphics[scale=0.4]{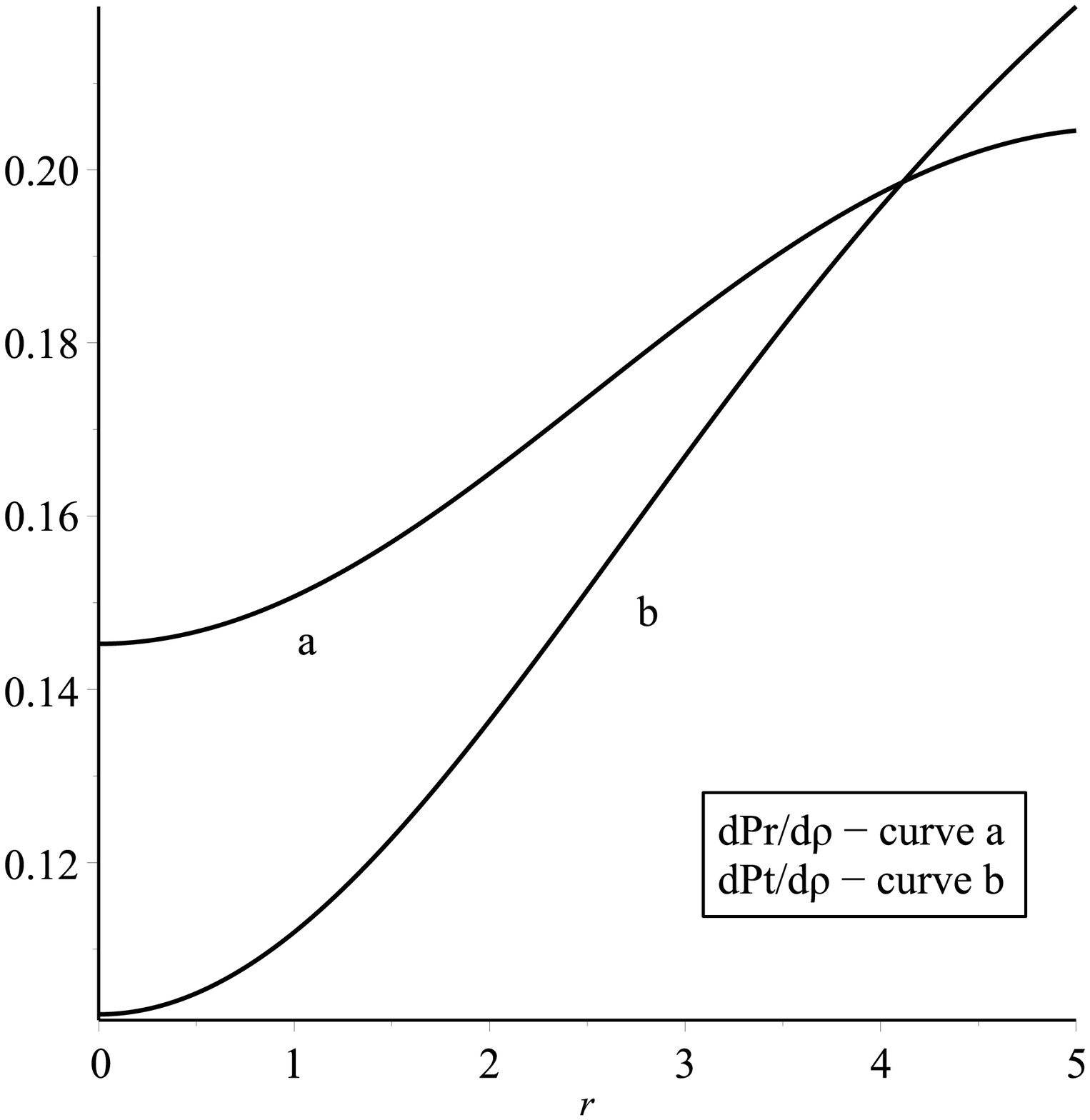}
\end{center}
\caption{Variation of radial and tangential pressure due to the density.}
\label{derivadasPder02}
\end{figure}

Now, we apply the matching conditions between the interior anisotropical solution and exterior Schwarzschild solution. It can be check that we have 3 unknown parameters,
namely $\{\alpha,\beta=\frac{A}{B}, C\}$, for chosen values of the mass $M_0$ and radius $R$. In particular, we will consider when these values are related with physical systems. The scaling constant $D$ can be set equal to one.

We notice that the values that the parameter $\beta$ shall take in order to obtain a physical solution depends on the values of $\alpha$, then for each value of $\alpha$ we have a range
\begin{eqnarray}
\beta_{min}(\alpha) \leq \beta \leq \beta_{max}(\alpha).
\end{eqnarray}
Using the continuity of the first fundamental form we found 
\begin{equation}
C=\frac{2M}{R^2(R-2M)}. \label{C}
\end{equation}
The exact FS analytic solution to the system of equations minimally deformed by an anisotropic source $\theta_{\mu\nu}$ is given by
\begin{eqnarray}
\label{pressrf}
\frac{8\pi \tilde{P}_r(r,\alpha,\beta)}{C}
&\!\!=\!\!&
(1-\alpha)H
\\ ,
\frac{8\pi \tilde{\rho}(r,\alpha,\beta)}{C}
&\!\!=\!\!&
\frac{8\pi \rho}{C} +\alpha\left[\frac{(\xi^2 +2)}{\xi^4}\frac{H}{(1+(\xi^2-1)H)}+\left(\frac{\xi^2-1}{\xi^3}\right)\frac{G(\xi)-2\xi H^2}{(1+(\xi^2-1)H)^2}\right]
\end{eqnarray}
\begin{eqnarray}
\label{pretanf}
\frac{8\pi \tilde{P}_t(r,\alpha,\beta)}{C}
&\!\!=\!\!&
H -\frac{\alpha}{2\xi^2(1+(\xi^2-1)H)}\left\lbrace (\xi^2-1)H\left(\frac{\xi^2+1}{\xi^2}+H(3\xi^2-1)\right) \right. \\ & + & \left.\left[\frac{H}{\xi^2}+\frac{1}{2}\left(\frac{\xi^2-1}{\xi}\right)\frac{G(\xi)-2\xi H^2}{(1+(\xi^2-1)H)}\right]\left(\left(\frac{\xi^2-1}{\xi}\right)\frac{d\nu}{d\xi}+2\right) \right\rbrace
\ .
\end{eqnarray}
where
\begin{eqnarray}
H&=&\frac{8\pi p}{C} , \\
G(\xi)&=& -\Big\{H\Big(\frac{\xi^2+1}{\xi}+\frac{1}{2}H\xi^3\Big)+\frac{2+\xi^2}{2\xi^3}\Big\}\\
\frac{d\nu}{d\xi}&=& \xi(1+\xi^2H).
\end{eqnarray}

We notice that we can recover the perfect fluid solutions when the value of the coupling constant is zero.

According to Eq.~(\ref{anisotropy}), the source $\theta_{\mu\nu}$ generates an anisotropy
given by
\begin{eqnarray}
\Pi(r,\alpha,\beta)
&\!\!=\!\!&
\alpha P -\frac{C}{8\pi}\frac{\alpha}{2\xi^2(1+(\xi^2-1)H)}\left\lbrace (\xi^2-1)H\left(\frac{\xi^2+1}{\xi^2}+H(3\xi^2-1)\right) \right. \\ & + & \left.\left[\frac{H}{\xi^2}+\frac{1}{2}\left(\frac{\xi^2-1}{\xi}\right)\frac{G(\xi)-2\xi H^2}{(1+(\xi^2-1)H)}\right]\left(\left(\frac{\xi^2-1}{\xi}\right)\frac{d\nu}{d\xi}+2\right) \right\rbrace
\ .
\end{eqnarray}

In resume, in order to find a realistic model the values for $M$, $r_s$ must be chosen. Then calculate $C$ using (\ref{C}), obtain $\beta$ with (\ref{A}) and $\xi_s$ by the definition. After this choice of values, we have to take care about the physical acceptability of the solution.

If we want to verify that this solution is physically acceptable, we have to check that the next conditions are satisfied
\begin{itemize}
\item $P_r$, $P_t$ and $\rho$ are positive and finite inside the distribution.
\item $\frac{dP_r}{dr}$, $\frac{dP_t}{dr}$ and $\frac{d\rho}{dr}$ are monotonically decreasing.
\item $\frac{P_r}{\rho}\leq 1$ \hspace{0.1cm}, \hspace{0.1cm} $\frac{P_t}{\rho}$ $\leq 1$.
\item $0<\frac{dP_r}{d\rho}<1$\hspace{0.1cm}, \hspace{0.1cm}$0<\frac{dP_t}{d\rho}<1$
\end{itemize}

In order to see which values of $\alpha$ will not lead to physical solutions, we check if the physical conditions are satisfied at the center of the distribution. We obtain that the values of the parameter $\alpha$ shall be restricted to
\begin{eqnarray}
\alpha\in [-1,0.5]-\{-0.02794,0.026023,0.0292,0.0865,0.311596\}.
\end{eqnarray}
Then, it can be found that the values of $\beta$ that satisfy the acceptability conditions in the center of the distribution are
\begin{eqnarray}
\beta\in\Big(\frac{\tan(1)-1}{1+\tan(1)}\, ,\, G_1(\alpha)\Big), &\mbox{  for  }& -1\leq\alpha <-0.02794 \\
 \beta\in\Big(\frac{\tan(1)-1}{1+\tan(1)}\, ,\, G_1(\alpha)\Big), &\mbox{  for  }& -0.02794< \alpha \leq 0 \\
  \beta\in\Big(\frac{\tan(1)-1}{1+\tan(1)}\, ,\, G_2(\alpha)\Big), &\mbox{  for  }& 0\leq\alpha\leq 0.01377\\
\beta\in\Big(\frac{\tan(1)-1}{1+\tan(1)}\, ,\, G_1(\alpha)\Big), &\mbox{  for  }& 0.01377\leq\alpha <0.026023 \\
 \beta\in\Big(\frac{\tan(1)-1}{1+\tan(1)}\, ,\, G_3(\alpha)\Big), &\mbox{  for  }& 0.026023\leq\alpha <0.0292 \\
  \beta\in\Big(\frac{\tan(1)-1}{1+\tan(1)}\, ,\, G_3(\alpha)\Big), &\mbox{  for  }& 0.0292<\alpha \leq 0.0865 
\end{eqnarray}
  
\begin{eqnarray} 
\beta\in\Big(-\infty \, , \,G_3(\alpha)\Big)\cup\Big(\frac{\tan(1)-1}{1+\tan(1)} \, ,\, \infty\Big) &\mbox{ for }& 0.0865<\alpha \leq 0.25, \\
\beta\in\Big(-\infty \, , \,G_4(\alpha)\Big)\cup\Big(\frac{\tan(1)-1}{1+\tan(1)} \, ,\, \infty\Big) &\mbox{ for }& 0.25<\alpha \leq 0.311596, \\
  \beta\in\Big(\frac{\tan(1)-1}{1+\tan(1)}\, ,\, G_3(\alpha)\Big), &\mbox{  for  }& 0.311596<\alpha \leq 0.5 
\end{eqnarray}
where
\begin{eqnarray}
G_1(\alpha)&=&\frac{-1}{2}\Big\{\frac{(25\alpha- 2)\sin(2)+2(4\alpha -9)\cos^2(1)+\sqrt{16\alpha^2 -572\alpha +21}-4\alpha +9}{2(\alpha -9)\sin(2)+(2-25\alpha)\cos^2(1)+25\alpha +3}\Big\}\\
G_2(\alpha) &=& \frac{-1}{2}\Big\{\frac{2(11\alpha- 1)\sin(2)+2(4\alpha -9)\cos^2(1)+\sqrt{16\alpha^2 -512\alpha +21}-4\alpha +9}{2(\alpha -9)\sin(2)+2(1-11\alpha)\cos^2(1)+22\alpha +3}\Big\} \\
G_3(\alpha)&=&  \frac{(2\alpha+1)+2(1-\alpha)\mbox{tan(1)}}{2(1-\alpha)-(1+2\alpha)\mbox{tan(1)}} \\
G_4(\alpha)&=&\frac{2-5\alpha +(1-\alpha)\mbox{tan(1)}}{1-\alpha -(5\alpha -2)\mbox{tan(1)}}
\end{eqnarray}

Physical solutions must live in this interval of values for the parameters $\alpha$ and $\beta$, but not all the parameters within this intervals are related to physical solutions, because the physical conditions must be satisfied for all values of $r$.

In fact, using maple, we prove that all the physical acceptability conditions are satisfied for $0.0865\leq\alpha<0.311596$, for all values of $r$, with $M=1$ and $R=5$. In particular, we show in the figures (\ref{prpt02})-(\ref{derivadasPder02}) the conditions for $\alpha=0.2$. 
\subsection{Solution~II: A condition for the density}
\label{s5.2}

Another restriction that can be imposed on the components of $\theta_{\mu\nu}$ in order to solve the system is 
\begin{equation}
\label{Dmimic}
\theta_0^{\,\,0}
=
\rho
\ .
\end{equation}

This constraint implies that the first order differential equation for $f(r)$ is
\begin{equation}
f'(r)+\frac{f(r)}{r}
=
-r\,k^2\,\rho
\ .
\end{equation}

It can be seen that the solution is given by
\begin{equation}
\label{Df}
f(r)
=-\frac{\xi^2-1}{\xi^2}+\frac{F}{\sqrt{\xi^2-1}}
\ ,
\end{equation}
where the density $\rho$ in Eq.~(\ref{tolmandensity}) has been used and $F$ is an integration constant. Then Eq.~(\ref{expectg}) yields
\begin{equation}
\label{D11}
e^{-\lambda(r)}
=
\xi^2
-\alpha\frac{\xi^2-1}{\xi^2}
\ .
\end{equation}
where we require finitness in the origin of the distribution, in consequence, F must be equal to zero.
\begin{figure}[h!]
\begin{center}
\includegraphics[scale=0.4]{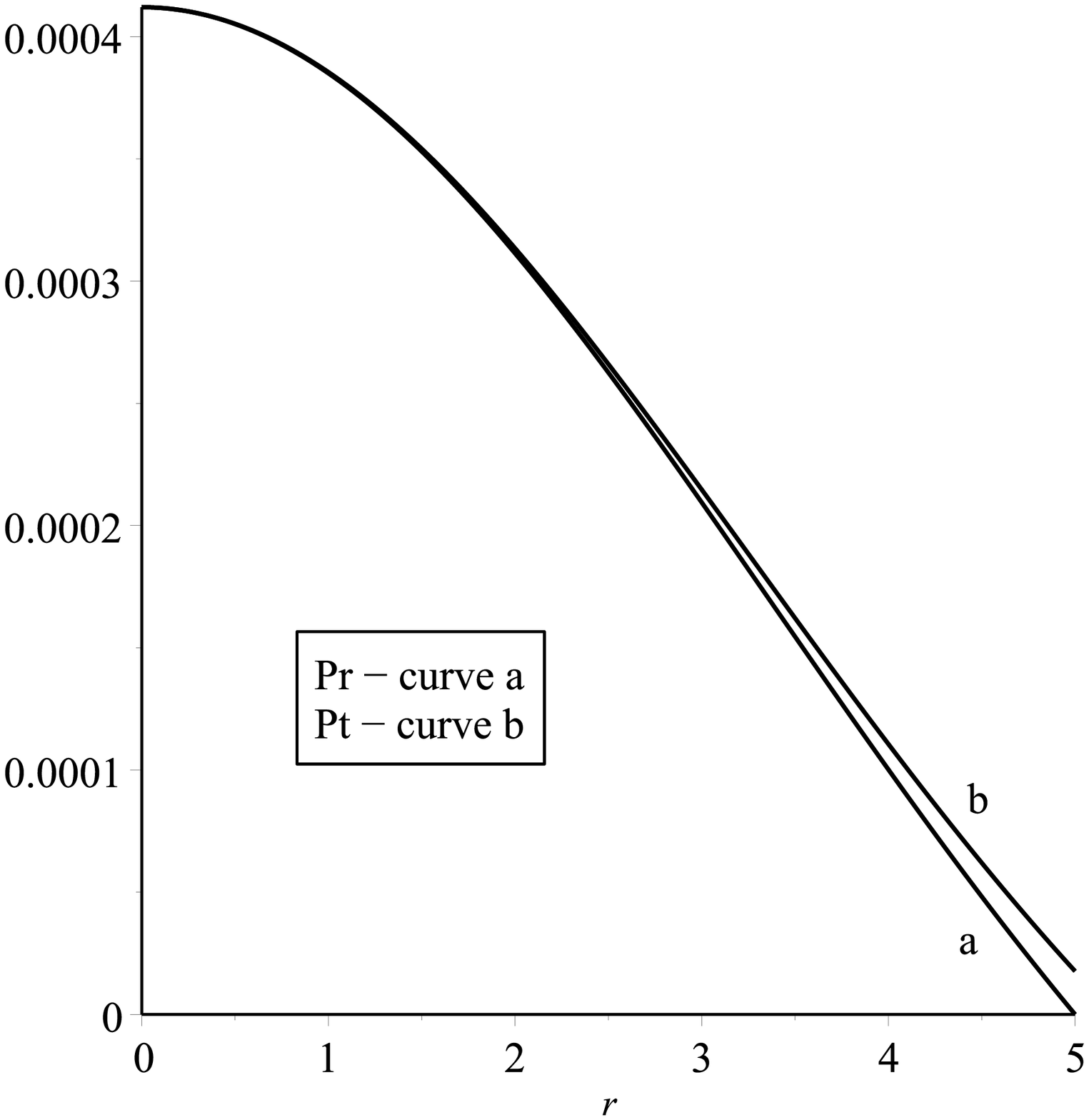}
\end{center}
\caption{ Radial and tangential pressure versus the radius}
\label{prpt}
\end{figure}
\begin{figure}[h!]
\begin{center}
\includegraphics[scale=0.4]{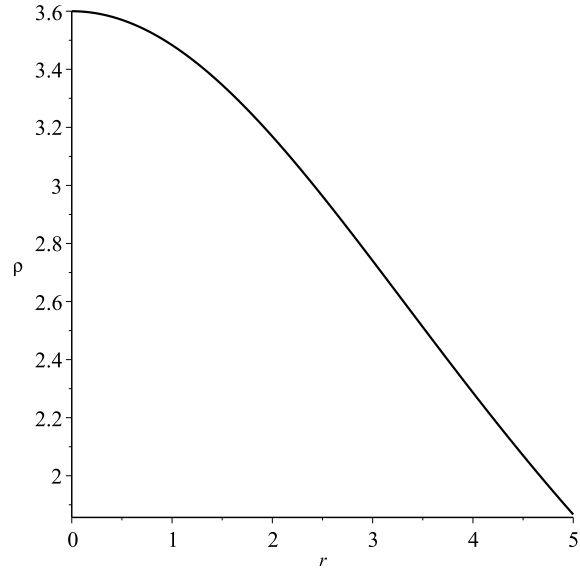}
\end{center}
\caption{ Density versus the radius}
\label{rhor}
\end{figure}
\begin{figure}[h!]
\begin{center}
\includegraphics[scale=0.4]{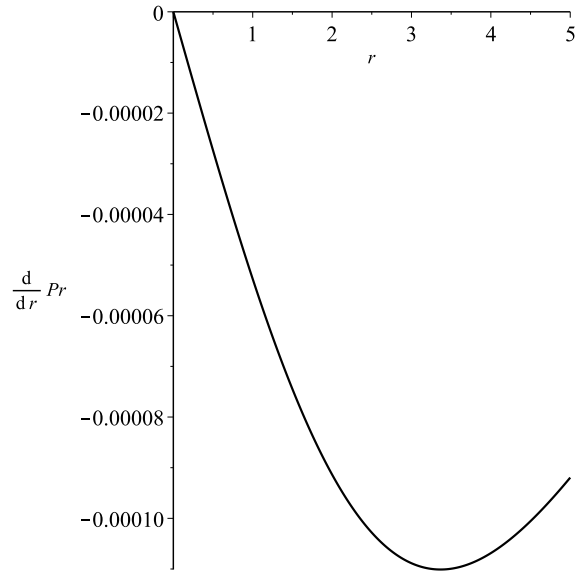}
\end{center}
\caption{Radial pressure gradient versus the radius}
\label{diffPr}
\end{figure}
\begin{figure}[h!]
\begin{center}
\includegraphics[scale=0.4]{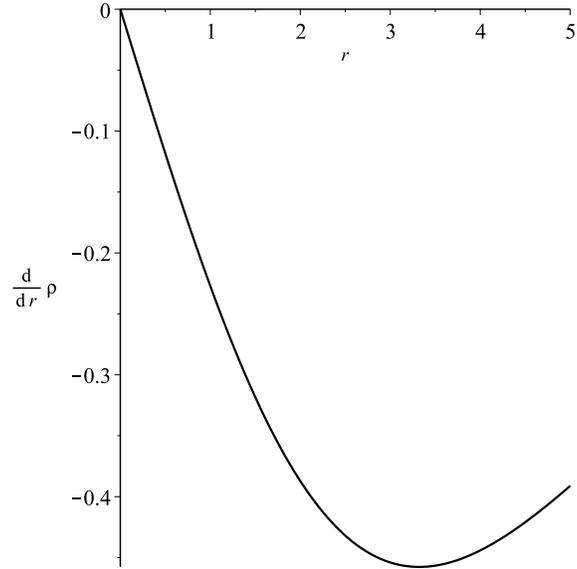}
\end{center}
\caption{Density gradient versus the radius}
\label{diffrho}
\end{figure}
\begin{figure}[h!]
\begin{center}
\includegraphics[scale=0.4]{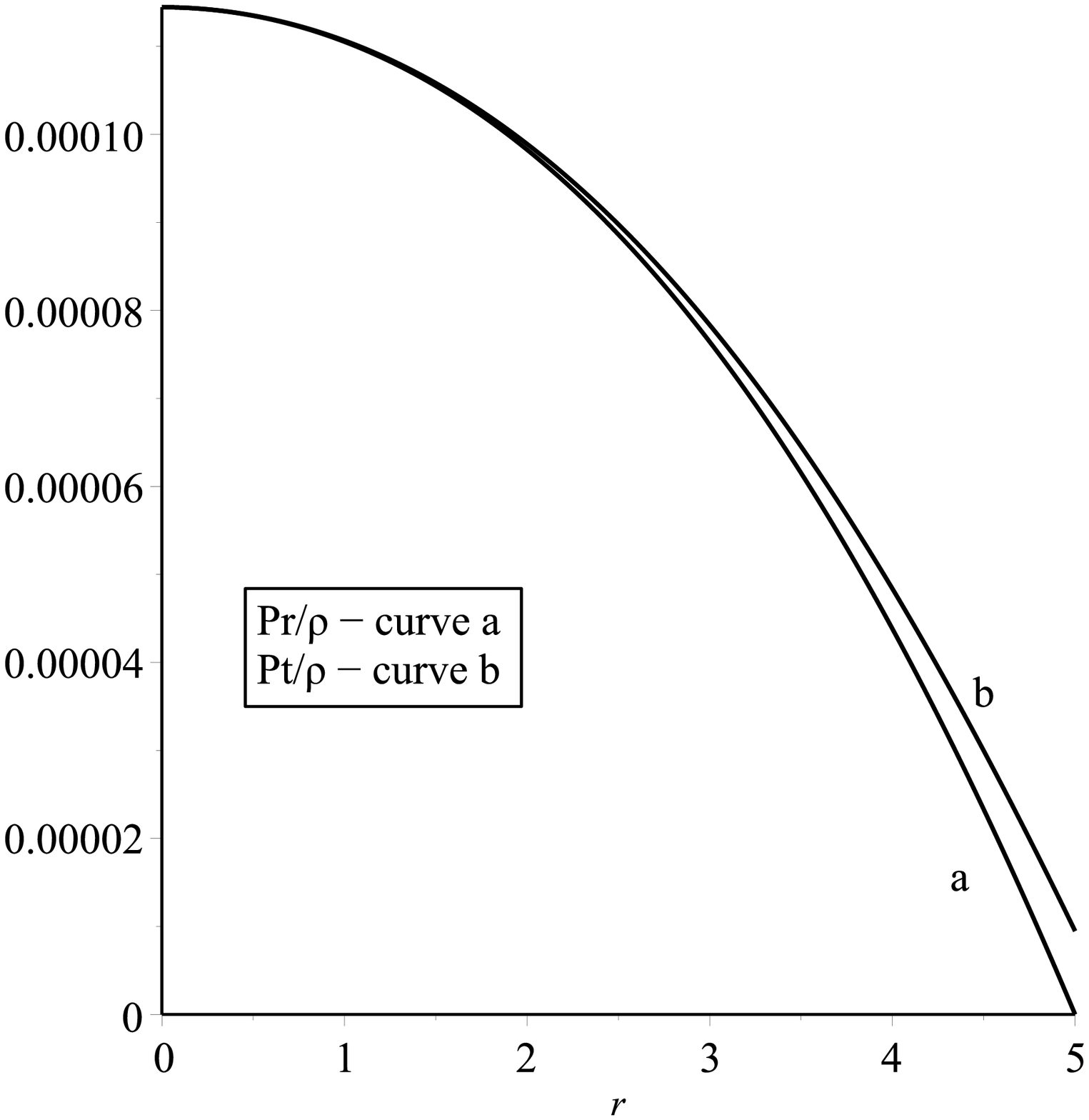}
\end{center}
\caption{Quotient of the radial and tangential pressure versus the radius}
\label{psrho}
\end{figure}   
\begin{figure}[h!]
\begin{center}
\includegraphics[scale=0.4]{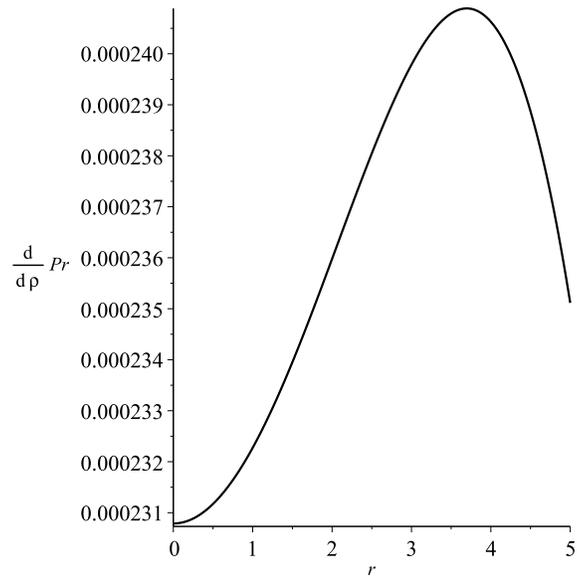}
\end{center}
\caption{Variaton of the radial pressure due to the density.}
\label{diffPrho}
\end{figure}
\begin{figure}[h!]
\begin{center}
\includegraphics[scale=0.4]{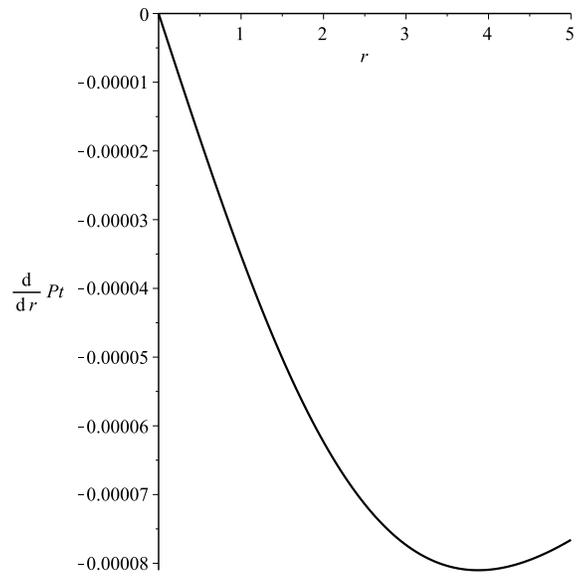}
\end{center}
\caption{Gradient of the tangential pressure versus the radius}
\label{dptt}
\end{figure}

After obtain $f^*$, the effective parameters can be written as
\begin{eqnarray}
\tilde{\rho}(r,\alpha) & = & (1+\alpha)\rho =\frac{C(1+\alpha)}{8\pi}\left(\frac{2+\xi^2}{\xi^4}\right) \\
\tilde{P_r}(r,\alpha,\beta) & = & p[1-\alpha(\xi^2-1)]-\frac{C\alpha}{8\pi} \\
\tilde{P_t}(r,\alpha,\beta) & = & p\left(1-\frac{3}{2}\alpha(\xi^2-1)\right)-\frac{C\alpha}{16\pi}\left(\frac{\xi^2-1}{\xi^2}\right)
\end{eqnarray}
with the anisotropy thus given by
\begin{equation}
\label{DPI}
\Pi(r,\alpha,\beta)
=
\alpha p\Big[-(\xi^2-1)-\frac{C\alpha}{8\pi}\Big(1-\frac{1}{2}\frac{(\xi^2-1)}{\xi^2}\Big)\Big]
\ .
\end{equation}
Also we found
\begin{eqnarray}
C & = & \frac{2M}{R^2(R(\alpha +1)-2M)} \label{cc}
\end{eqnarray}
due to the coupling conditions.

Following the same steps as before, we can get another new analytical solution. If we choose $R=5$, $M=1$,$\alpha=0.2$ y $D=1$, we obtain then $C\approx 0.02$  y $\beta\approx 1.109$. The figures (\ref{prpt})-(\ref{dptt}) shows that the physical conditions are satisfied for this specific solution.

Equivalently to the mimic constraint for pressure, it can be shown that any other possible physical solution for the selected values of $M$ and $R$, if exists, it shall be related to the parameters within the interval
\begin{eqnarray}
\alpha\in \Big[-0.6,4.588\Big)-\{0.397\}.
\end{eqnarray}
The associated values of $\beta$ are
\begin{eqnarray}
\beta\in\Big[\frac{1-\alpha-(1+\alpha)\mbox{tan(1)}}{(\alpha-1)\mbox{tan(1)}-1-\alpha} \, , \, \frac{(1+2\alpha)+2(1+\alpha)\mbox{tan(1))}}{2(1+\alpha)-(1+2\alpha)\mbox{tan(1)}} \Big]
\end{eqnarray}
for $-0.6\leq\alpha<0.397$
\begin{eqnarray}
\beta\in\Big(-\infty \, ,\, \frac{(1+2\alpha)+2(1+\alpha)\mbox{tan(1)}}{2(1+\alpha)-(1+2\alpha)\mbox{tan(1)}}\Big]\cup\Big[\frac{1-\alpha-(1+\alpha)\mbox{tan(1)}}{(\alpha-1)\mbox{tan(1)}-1-\alpha}\, , \, \infty\Big)
\end{eqnarray}
for $0.397<\alpha<4.588$.

Outside of this intervals, the conditions for a solution to be physical are not satisfied, but not all the possible choices within this interval will lead to physical solutions because the other physical conditions could not be satisfied.

\section{Conclusions}

There are evident advantages in the MGD method as an efficient approach to decouple gravitational sources in Einstein's equations. MGD approach splits the usual system of equations in two, the first one identical to the isolated perfect fluid case. The second one, similar to Einstein's equations, associated with the gravitational sources whose interaction with the perfect fluid is purely gravitational. 

We were able to extend the known FS solution to the anisotropical domain using the MGD method. This was done for two different conditions on the gravitational source. We found new anisotropical solutions for the mimic constraint for pressure and density. The chosen values for the parameters like the ratio at the surface $R=5$ and the mass $M=1$ are the typical values related to systems with high density mass. This is important because is not easy to find analytical solutions of Einstein's equations with physical relevance, which eventually could ease the study of some properties of self-gravitating compact realistic systems.

Moreover, we found an interval of values for the parameters $\alpha$ and $\beta$, where all physical solutions shall dwell. This was done by the imposition of the physical conditions, at least, in the center of the distribution, for both constraints. For the mimic constraint for the pressure, we were able to find an interval of $\alpha$ where all physical acceptability conditions are satisfied for all values of $r$. Except by the latter, all other intervals for both constraints must improve from the consideration of the physical conditions from $r=0$ to $r=R$. 
\label{s8}

\setcounter{equation}{0}
\section{Acknowledgements}

P. Le\'on and C. Las Heras want to say thanks for the Financial
help received by the Projects ANT1756 and ANT1755
of the Universidad de Antofagasta, respectively.
\end{document}